\documentclass[a4paper,12pt]{article}

\usepackage{fullpage}
\usepackage[T1]{fontenc}

\usepackage{amssymb,amsmath}

\usepackage{graphicx}

\usepackage{amsmath,amsfonts,epsfig,mathrsfs,todonotes,yfonts,cite}  
\newcommand{\nn}{\nonumber}

\newcommand{\re}[1] {(\ref{#1})}
\newcommand{\rd}[1]{{\color{red}{#1}}}

\newcommand{\al}{\alpha} 
\newcommand{\be}{\beta}

\newcommand{\de}{\delta}

\newcommand{\beq}{\begin{equation}}
\newcommand{\eeq}{\end{equation}}

\newcommand{\ber}{\begin{eqnarray}}
\newcommand{\eer}[1]{\label{#1}\end{eqnarray}}
\newcommand{\eero}{\end{eqnarray}}

\newcommand{\half}{{\textstyle{\frac12}}}

\newcommand{\halfp}{{\textstyle{\frac1{2\pi}}}}

\newcommand{\pa}[1]{\partial_{#1}}

\begin{document}

\setcounter{footnote}{0}

\baselineskip 6 mm

\begin{titlepage}
	\thispagestyle{empty}
	\begin{flushright}
		
	\end{flushright}
\rightline{{UUITP-15/26}}

	\vspace{35pt}
	
	\begin{center}
	    {{\Large\bf The conformal null string in $d+2$ and $d$ dimensions} }

		\vspace{50pt}
		
		{Ulf Lindstr\"om$^{a,b,c}$}   
		
		\vspace{25pt}

		\vspace{15pt}

        $^{(a)}${\it Department of Physics and Astronomy, Division of Theoretical Physics, Uppsala University, \\ 
        Box 516, SE-75120 Uppsala, Sweden}
		
		\vspace{15pt}

        $^{(b)}${\it Center for Geometry and Physics, Uppsala University, Box 480, SE-75106 Uppsala, Sweden}
		
		\vspace{15pt}      
		
		$^{(c)}${\it Physics Division, National Technical University of Athens \\
        15780 Zografou Campus, Athens, Greece}
  \end{center}
		
		\vspace{40pt}
		
		{ABSTRACT} \\
		
\noindent	In \cite{Lindstrom:2026quz} it is pointed out how the tensionless string with a gauged scale symmetry discussed in the recent articles, \cite{Sheikh-Jabbari:2026cnj,Sheikh-Jabbari:2026vqh,Sheikh-Jabbari:2026tpf}, is a reduction of the conformal string \cite{Gustafsson:1994kr} to Minkowski space. Here we corroborate this by choices of slices in Dirac $d+2$  dimensional conformal space. We perform a Dirac reduction of the model and its  algebra of constraints and see how they map to the constraints in $d$ dimensions, including how the semidirect product of the Virasoro algebra with a su(1,1) Kac-Moody algebra becomes the Corrollian-Weyl symmetry in  $d$ dimensions.

\vspace{\fill}

	\noindent{\footnotesize{Email: ulf.lindstrom@physics.uu.se
 }}

\vspace{10pt}

\bigskip

\end{titlepage}
\section{Introduction}

Null strings were introduced in \cite{Schild:1976vq} and later rediscovered in a different guise in \cite{Karlhede:1986wb}.
Their quantisation has been an issue for a long time and depends on its symmetries and constraints as well as the choice of operator ordering and vacuum. It has been discussed in many papers, most recently in \cite{Duary:2026rlo}. For that reason the contribution of the interesting  recent paper \cite{Sheikh-Jabbari:2026cnj} becomes important. The $d$ dimensional model in these papers is constructed following the same pattern used for the conformal null string in Dirac's $d+2$ dimensional conformal space\footnote{Dirac type embeddings has recently found extensive applications  in superspace. See \cite{Koning:2026rkc} and references therein. } more than thirty years ago \cite{Gustafsson:1994kr}.  The constraints can also be seen to arrive in the same way. In fact solving the light cone constraint for the conformal string gives this $d$ dimensional model as will be shown here. 

For the $d$ dimensional model in \cite{Sheikh-Jabbari:2026cnj,Sheikh-Jabbari:2026vqh,Sheikh-Jabbari:2026tpf}  focus is on a local scale transformation of the spacetime coordinates compensated by a transformation of the world sheet vector densities, and brings out a Carrollian-Weyl symmetry of the theory. Constant scaling has previously been included in the conformal group of the ambient spacetime \cite{Isberg:1992ia}, whereas the local version was studied in  $d+2$ dimensions in \cite{Gustafsson:1994kr} where the scale generator is part of the constraints. The full constraint algebra  was identified as a semidirect product of the Virasoro algebra with a $su(1,1)$  Kac-Moody algebra. 

 We feel that it is important to detail the relation to the $d$ dimensional approach of the recent papers \cite{Sheikh-Jabbari:2026cnj,Sheikh-Jabbari:2026vqh,Sheikh-Jabbari:2026tpf} and the present paper is dedicated  to that endeavour. 

 Whereas solving the embedding constraint for the conformal string produces the gauged $d$ dimensional model, it does not directly reduce all the constrains and their algebra. We therefore perform a Dirac reduction of the higher dimensional model to obtain the constraints in $d$ dimension and map the semidirect product of the Virasoro algebra with $su(1,1)$  Kac-Moody algebra to the Carrollian-Weyl symmetry.

\section{The conformal string in Dirac space}
The ILST formulation of the zero tension limit of the bosonic string, first derived as the tensionless limit of the Nambu-Goto string\footnote{It also follows from the the tensionless limit of the "Polyakov" string} in \cite{Lindstrom:1990qb, Lindstrom:1990ar} and discussed in detail in, e.g., \cite{Isberg:1993av}, reads
\beq\label{act}
S=\int d^2\xi V^\al V^\be\pa{\al} X^m\pa{\be} X^n\eta_{mn}
\eeq
where $X^m$ are coordinates of the ambient $d$ dimensional flat space time,  
$\xi^\al=(\tau.\sigma)$  are coordinates on the world  sheet, and $V^\al$ are two $2$-dimensional worldsheet vector densities. The classical symmetries are conformal symmetries of the ambient space time and worldsheet diffeomorphisms. Rigid scale invariance is part of the conformal invariance, but as we shall see there is also an important local scale invariance.

The conformal string\ constructed in  \cite{Gustafsson:1994kr}  is a direct application of the same construction in Dirac's $d+2$ dimensional conformal space with the motion restricted to the higher dimensional light cone\footnote{This is the embedding used in finding an alternative version of the zero tension limit of the bosonic string in \cite{Karlhede:1986wb}. For further possible formulations of tensionless strings and branes see \cite{Hassani:1994rf, Lindstrom:1997uj}~.
}:
\beq\label{Hamact}
S_c=\int d^2\xi \big(V^\al V^\be {\cal D}_{\al} \hat X^M{\cal D}_{\be} \hat X^N\hat \eta_{MN}+\Phi \hat X^2\big)~.
\eeq
Here $\hat X^M$ are coordinates in the higher dimensional space of signature $(-+++...+-)$ with metric  
\beq
\left( \begin{array}{lll}
\eta_{mn}&0&0\\
0&1&0\\
0&0&-1
\end{array}
\right)
\eeq
and $\Phi $ is a Lagrange multiplier upholding the lightcone (embedding) condition.  In addition the covariant derivatives are
\ber
{\cal D}_{\al} =\pa{\al}+W_\al
\eer
where $W_\al$ is the gauge field for scale transformations
\begin{align}\nn
&\de X^M=\lambda X^M\\\nn
&\de V^\al=-\lambda V^\al\\\nn
&\de W_\al=-\pa{\al}\lambda\\
&\de\Phi=-2\lambda \Phi~.
\end{align}
This is how the local scale transformations enter the model. Additional symmetries  are 
$2$-d diffeomorphisms, rotations in the higher dimensional space  and some ``additional'' symmetries. 

\section{Choosing a physical slice}
To see the conformal string in Minkowski space we need to choose a physical slice, a 
parametrisation of the the light cone that introduces the Minkowski coordinates. Here we illustrate the procedure in two ways. In both cases we make a coordinate transformation to change the metric to
\beq\label{met2}
\left( \begin{array}{lll}
\eta_{mn}&0&0\\
0&0&\half\\
0&\half&0
\end{array}
\right)
\eeq
which implies 
\beq\label{LC1}
\hat X^2=\hat X^m\hat X^n\eta_{mn}+\hat X^+\hat X^-~,
\eeq
where $\hat X^\pm=\hat X^d\pm \hat X^{(d+1)}$~.

\bigskip

\noindent
i) {\em Slice choice \`a la Dirac}: 

Choose $\hat X^M=(X^m,1,-X^2)$, which is fine due to the projective invariance of the light cone condition. Then the light cone condition is satisfied and
\beq
{\cal D}_{\al} \hat X^M{\cal D}_{\be} \hat X^N\hat \eta_{MN}={\cal D}_{\al} \hat X^m{\cal D}_{\be} \hat X^n \eta_{mn}-{\cal D}_{\al}1{\cal D}_{\be} X^2={\cal D}_{\al} \hat X^m{\cal D}_{\be} \hat X^n \eta_{mn}~.
\eeq
Then
\beq\label{actS}
S_c\to\int d^2\xi V^\al V^\be{\cal D}_{\al} \hat X^m{\cal D}_{\be} \hat X^n \eta_{mn}
\eeq
which is the form of the action employed in \cite{Sheikh-Jabbari:2026cnj,Sheikh-Jabbari:2026vqh,Sheikh-Jabbari:2026tpf}. Notice that the choice $\hat X^+=1$ breaks the scale symmetry.
\bigskip

\noindent
ii) {\em Following Barut} \cite{Barut}: 

Choose $\hat X^M=(\varphi X^m,\varphi,\chi)$.  Then the light cone condition is satisfied with $\chi=-\varphi X^2$. Inserting the conditions we eventually find that 
\beq
{\cal D}_{\al} \hat X^M{\cal D}_{\be} \hat X^N\hat \eta_{MN}=\varphi^2{\cal D}_{\al} \hat X^m{\cal D}_{\be} \hat X^n \eta_{mn}~,
\eeq
so that
\beq\label{act}
S_c\to\int d^2\xi \varphi^2V^\al V^\be{\cal D}_{\al} \hat X^m{\cal D}_{\be} \hat X^n \eta_{mn}~.
\eeq
Redefining $\varphi V^\al\to V^\al$ we again find the action \re{actS}.

\section{Constraints and algebra}

The momentum $\hat P_M$ is the derivative of the Lagrangian in \re{Hamact} with respect to $\pa{\tau}\hat X$:
\beq\label{moment}
\hat P_M=2V^\al{\cal D}_{\al} \hat X_N\frac {\partial V^\be{\cal D}_{\be} \hat X^N}{\partial\pa{\tau}\hat X^M}=2V^\tau V^\al{\cal D}_{\al} \hat X_M~\to~V^\al{\cal D}_{\al} \hat X_M~,
\eeq 
where we rescaled the momentum at the end.
Since we are dealing with a ($2$-dim) diffeomorphism invariant theory, the Hamiltonian will be the sum of the constraints, each multiplied by Lagrange multipliers. In the Lagrangian formulation these multipliers are $V, W$ and $\Phi$.  We find the set of four constraints for the conformal string
\begin{align}\label{consts1}\nn
&\phi^T=\hat P_M\dot {\hat X}^M ~,~~\phi^{0}=\hat P_M\hat X^M~,~~\phi^{1}=\hat X^2~,~~ \phi^L=\hat P_M\hat X'^M\\\nn
&~~~~~~~~~~~~~~~~~~~~~~~~~~~~~~\Leftrightarrow\\
& \phi^{-1}=\hat P_M\hat P^M~,~~\phi^{0}=\hat P_M\hat X^M~,~~\phi^{1}=\hat X^2~,~~ \phi^L=\hat P_M\hat X'^M
\end{align}
where $\hat P_N$ is the momentum conjugate to $\hat X^M$ and we used $V^\tau\phi^T=\phi^{-1}-V^\sigma \phi^L$ in the second line, which is the form of the constraints employed in \cite{Gustafsson:1994kr}.
 We may investigate the constrain algebra using Poisson brackets, such as
\beq
\{\hat X^M(\sigma),\hat P_N(\sigma'\}=\de^M_N\de(\sigma-\sigma')~.
\eeq{}
It is shown in \cite{Gustafsson:1994kr} that these constraints form a primary system and thus serve as a good starting point for quantisation.

We shall reduce the constraints in the second line of \re{consts1} to $d$ dimensions and compare to the Minkowski space formulation of \cite{Sheikh-Jabbari:2026vqh}. \\
In a preliminary analysis  we make use of the $2$-dimensional diffeomorphism gauge choice\footnote{In \cite{Gustafsson:1994kr}  the less restrictive choice $V^\al = (E^{-\half}(\tau),0)$ is made to show that the conformal string represents a collection of conformal particles \cite{Marnelius:1978fs}
.} $V^\al = (1,0)$, the temporal gauge  introduced in 
\cite{Lindstrom:1990qb}. 
Using the metric \re{met2} we have, with $\hat P_M=\{P_m,P_+,P_-\}$,
\begin{align}\label{consts}\nn
&\phi^{-1}=P_mP^m+ P_+P_-=0\\\nn
&\phi^{0}=P_m\hat X^m+\half P_+\hat X^-+\half P_-\hat X^+=0\\\nn
&\phi^{1}=\hat X^m\hat X_m+\hat X^+\hat X^-=0\\
&\phi^L=P_m\hat X'^m+\half\hat X'^-P_++\half P_-\hat X'^+=0~,
\end{align}
Choosing the  slice in i); $\hat X^M=(X^m, 1, \hat X^-)$ solves the constraint $\phi^{1}$, when $\hat X^-=-X^2$. In temporal gauge, the momenta now read 
\begin{align}\nn
&P_m=\dot{\hat X}_m+W_\tau \hat X_m=\dot X_m+W_\tau X_m\\\nn
&P_+=\dot{\hat X}_++W_\tau \hat X_+ =W_\tau \hat X_+\\
&P_-=\dot{\hat X}_-+W_\tau \hat X_-=\dot\chi +W_\tau \chi =- 2X_m\dot X^m-W_\tau X^2 ~,
\end{align}

Using this we find 
\begin{align}\label{tempcon}\nn
&\phi^{-1}:~~\hat P_M\hat P^M= P_mP^m+P_+P_-= \dot X_m\dot X^m\\
&\phi^L: ~~~\hat P_M\hat X'^M=P_mX'^m+\half P_+\hat X'^-+\half P_-\hat X'^+= \dot X_m X'^m
\end{align}
These are two of the $d$ dimensional constraints for the conformal string. Missing is the constraint  $P_mX^m$. We might have expected that to arise from $\phi ^0: P_M\hat X^M$, but this scale constraint is identically zero for our coordinatisation of $\phi^1: \hat X^2=0$, which breakes scale invariance. %(Note that $\phi^0$ and $\phi^1$ are conjugate and form a first class gauge system.) 

So we proceed differently, do not use the temporal gauge and instead of solving $\phi^1$ directly, we follow Dirac and use second class constraints to eliminate degrees of  freedom. Impose the constraint  
\beq
\chi: \hat X^+-1=0~.
\eeq
This is a second class constraint since $[\phi^0,\chi]=-\hat X^+ \ne 0.$ We shall need to pair it with another second class constraint which reduces to $\phi^1$ on the constraint surface and which does not break scale invariance. A good choice for this is
\beq
\tilde\phi^1:= \frac {\hat X^2}{(\hat X^+)^2}
\eeq
which is scale invariant everywhere since $\hat X^M \to e^\lambda \hat X^M$ implies that $\tilde\phi^1$ is invariant. It is also a second class constraint since $[\tilde\phi^1,\phi^{-1}]$ and $[\tilde\phi^1,\phi^{L}]$  are non vanishing. Finally $[\tilde\phi^1,\chi]=0$ ensures that the Dirac brackets become Poisson brackets on the constraint surface,
and  that fixing the scale gauge via $\chi $ does not interfere with the geometry of the conformal string shell condition ${\phi }^{1}$.

We may now strongly set the two second class constraints to zero to eliminate two degrees of freedom, one position and one momentum variable\footnote{An additional pair will be removed from solving the light cone condition, see below.}. We may then legally use $\hat X^M=(X^m, 1, -X^2)$ without breaking scale invariance, since the constraints $\phi^{-1}, \phi^{0}$ and $\phi^L$ still  remain. This reduces the Hamiltonian according to
\beq
H=\lambda_i\phi^i \to H_{red}=\lambda_{-1}\phi^{-1}+\lambda_{0}\phi^{0}+\lambda_{L}\phi^{L}~,
\eeq
and forces us to use Dirac instead of Poisson brackets when evaluating the remaining $\phi^i$s  on the restricted phase space.

On the restricted phase space we solve $\phi^0$ for $P_-$
\beq\label{pmin}
P_-=-2PX+ P_+X^2
\eeq
and $\phi^L$ for $P_+$
\beq\label{pplus}
P_+=\frac{PX'}{XX'}
\eeq
When inserted in $\phi^{-1}$ we then have the nonlinear constraint
\beq
P^2+\frac{PX'}{XX'}(-PX+\half X^2\frac{PX'}{XX'})
\eeq 
which we write as 
\beq
(XX')^2P^2+
{PX'}(-2(PX)(XX')+X^2(PX'))~.
\eeq
This is one single constraint on the reduced phase space. But we started from four first class constrains and used one of them to remove one variable. The remaining three are still first class (with respect to the Dirac bracket) and should reduce to three first class constraints. This forces the split up of the one constraint into the following three first class ones
\footnote{Formally there is also a branch without the $PX$ constraint which leads to model with only two constraints and lacking scale symmetry, but this is not in keeping with the expectation of three first class constraints.}
\ber\label{const2}
{\cal C}_1=P^2~,~~ {\cal C}_2=PX'~,~\makebox{and} ~~{\cal C}_3=PX~,
\eer{}
as anticipated from the temporal gauge results \re{tempcon}.

Note that we have eliminated two coordinates via the constraint $\chi$ and the solution of  $\phi^1$. The corresponding momenta are eliminated via \re{pmin} and \re{pplus} where $P_\pm$ are expressed completely in $d$-dimensional variables.

For background on constraints and Hamiltonian reductions, see \cite{Dirac:1950pj} and \cite{Henneaux:1994lbw}.

\section{The symmetry algebras}

{\bf The Virasoro-Kac-Moody symmetry in $d+2$}\\

The algebra of constraints in the conformal string, $\mathbb{C}_{d+2}$ , is a semi direct product of the Virasoro algebra with an $su(1,1)$ Kac-Moody algebra. Starting from the constraints in \re{consts}, define generators as 
\begin{align}
&L_m=\halfp \int _0^{2\pi}d\sigma e^{im\sigma}\phi^L(\sigma)~,~~~~
T^a_m=\halfp \int _0^{2\pi}d\sigma e^{im\sigma}\phi^a(\sigma)~,
\end{align}
where $a=\{-1,0,1\}$ and $m,n,...$ denote Fourier components, not spacetime indices~.

Then the {\em Virasoro subalgebra} is defined by
\beq
[L_m,L_n]=(m-n)L_{m+n}~,
\eeq 
and its action on the Kac-Moody currents is
\beq
[L_m,T^a_n]=(m(1-a)-n)T^a_{m+n}~.
\eeq
Under world sheet diffeomorphisms $T^a_m$ transforms as a primary field with conformal weight $\Delta_a=1-a.$ 

The {\em $su(1,1)$ Kac-Moody algebra} is
\beq
[T^0_m,T^{\pm 1}_n]=\mp 2 T^{\pm 1}_{m+n}~,~~~~[T^{-1}_m,T^{1}_n]=-4 T^{0}_{m+n}~.
\eeq
\bigskip

{\bf The Corrollian-Weyl algebra in $d$}\\

Starting from the constraints \re{const2}, $\mathbb{C}_d$, define  generators as
\beq
M_m=\halfp \int _0^{2\pi}d\sigma e^{im\sigma}{\cal C}_1(\sigma)~,~~~
L_m=\halfp \int _0^{2\pi}d\sigma e^{im\sigma}{\cal C}_2(\sigma)~,~~~
S_m=\halfp \int _0^{2\pi}d\sigma e^{im\sigma}{\cal C}_1(\sigma)~.
\eeq
Then the Corrollian-Weyl algebra is
\begin{align}\nn
&[L_m,L_n]=(m-n)L_{m+n}~,~~~[L_m,M_n]=(m-n)M_{m+n}\\\nn
&[L_m,S_n]=-nS_{m+n}~,~~~[S_m,M_n]=-2M_{m+n}\\
&[M_m,M_n]=0~,~~~[S_m,S_n]=0~.
\end{align}
The first line lists the Virasoro algebra and the supertranslations and the second line deals with the Weyl rescaling current.
\bigskip

{\bf $\mathbb{C}_{d+2}$ to  $\mathbb{C}_{d}$ via the Dirac reduction}\\

The Dirac reduction acts as a algebraic truncation mapping the higher-dimensional \({su}(1,1) \rtimes \text{Virasoro}$ structure to the Carrollian Weyl algebra:

\begin{itemize}

\item The worldsheet diffeomorphism generator $L_{n}$ maps directly down since $\phi^{L} \to 
{\cal C}_{2}$.

\item The ${su}(1,1)$ generator \(T_{n}^{-1}\) reduces directly to \(M_{n}\) (the supertranslations).
\item The neutral \({su}(1,1)\) generator \(T_{n}^{0}\) reduces to \(-S_{n}\) (from the scaling constraint).
\item The generator \(T_{n}^{1}\) is frozen into a constant c-number background state by the second-class gauge condition.

\end{itemize}

\section{Summary}

We have shown how the conformal string defined in $d+2$ Dirac space reduces along with its constraints and their algebra to the conformal string i $d$ dimensions, which is the subject of recent studies.  
\bigskip
 
\noindent
{\bf Acknowledgements}:\\
Extensive discussions with Bo Sundborg and 
correspondence with M. M. Sheikh-Jabbari and H. Yavartanoo are gratefully acknowledged.

\end{document}